\documentclass[aps,prl,amsmath,amssymb,twocolumn,showpacs,superscriptaddress,final,floatfix]{revtex4-1}
\usepackage{graphicx}	
\begin{document}

\title{Classification of Collective Modes in a Charge Density Wave by Momentum-Dependent Modulation of the Electronic Band Structure}

% ***** AUTHORS
\author{D. Leuenberger}
\affiliation{Stanford Institute for Materials and Energy Sciences, SLAC National Accelerator Laboratory, 2575 Sand Hill Road, Menlo Park, CA 94025, USA}
\affiliation{Geballe Laboratory for Advanced Materials, Department of Applied Physics, Stanford University, Stanford, CA 94305, USA}
\author{J.~A. Sobota}
\affiliation{Stanford Institute for Materials and Energy Sciences, SLAC National Accelerator Laboratory, 2575 Sand Hill Road, Menlo Park, CA 94025, USA}
\affiliation{Geballe Laboratory for Advanced Materials, Department of Applied Physics, Stanford University, Stanford, CA 94305, USA}
\affiliation{Lawrence Berkeley National Lab, 1 Cyclotron Road, Berkeley, CA 94720, USA}
\author{S.-L. Yang}
\affiliation{Stanford Institute for Materials and Energy Sciences, SLAC National Accelerator Laboratory, 2575 Sand Hill Road, Menlo Park, CA 94025, USA}
\affiliation{Geballe Laboratory for Advanced Materials, Department of Applied Physics, Stanford University, Stanford, CA 94305, USA}
\affiliation{Department of Physics, Stanford University, Stanford, CA 94305, USA}
\author{A.~F. Kemper}
\affiliation{Lawrence Berkeley National Lab, 1 Cyclotron Road, Berkeley, CA 94720, USA}
\author{P.~Giraldo-Gallo }
\affiliation{Stanford Institute for Materials and Energy Sciences, SLAC National Accelerator Laboratory, 2575 Sand Hill Road, Menlo Park, CA 94025, USA}
\affiliation{Geballe Laboratory for Advanced Materials, Department of Applied Physics, Stanford University, Stanford, CA 94305, USA}
\author{R.~G. Moore}
\affiliation{Stanford Institute for Materials and Energy Sciences, SLAC National Accelerator Laboratory, 2575 Sand Hill Road, Menlo Park, CA 94025, USA}
\affiliation{Geballe Laboratory for Advanced Materials, Department of Applied Physics, Stanford University, Stanford, CA 94305, USA}
\author{I.~R. Fisher}
\affiliation{Stanford Institute for Materials and Energy Sciences, SLAC National Accelerator Laboratory, 2575 Sand Hill Road, Menlo Park, CA 94025, USA}
\affiliation{Geballe Laboratory for Advanced Materials, Department of Applied Physics, Stanford University, Stanford, CA 94305, USA}
\author{P.~S. Kirchmann}
\email{kirchman@stanford.edu}
\affiliation{Stanford Institute for Materials and Energy Sciences, SLAC National Accelerator Laboratory, 2575 Sand Hill Road, Menlo Park, CA 94025, USA}
\author{T.~P. Devereaux}
\affiliation{Stanford Institute for Materials and Energy Sciences, SLAC National Accelerator Laboratory, 2575 Sand Hill Road, Menlo Park, CA 94025, USA}
\affiliation{Geballe Laboratory for Advanced Materials, Department of Applied Physics, Stanford University, Stanford, CA 94305, USA}
\author{Z.-X. Shen}
\email{zxshen@stanford.edu}
\affiliation{Stanford Institute for Materials and Energy Sciences, SLAC National Accelerator Laboratory, 2575 Sand Hill Road, Menlo Park, CA 94025, USA}
\affiliation{Geballe Laboratory for Advanced Materials, Department of Applied Physics, Stanford University, Stanford, CA 94305, USA}
\affiliation{Department of Physics, Stanford University, Stanford, CA 94305, USA}

% ***** DATE
\date{\today}

% ***** ABSTRACT
\begin{abstract}
We present time- and angle-resolved photoemission spectroscopy measurements on the charge density wave system CeTe$_{3}$. Optical excitation transiently populates the unoccupied band structure and reveals a gap size of $2\Delta =0.59$~eV. The occupied Te-$5p$ band dispersion is coherently modified by three modes at $\Omega_1 = 2.2$~THz, $\Omega_2 = 2.7$~THz and $\Omega_3 = 3$~THz. All three modes lead to small rigid energy shifts whereas $\Delta$ is only affected by $\Omega_1$ and $\Omega_2$. Their spatial polarization is analyzed by fits of a transient model dispersion and DFT frozen phonon calculations. We conclude that the modes $\Omega_1$ and $\Omega_2$ result from in-plane ionic lattice motions, which modulate the charge order, and that $\Omega_3$ originates from a generic out-of-plane A$_\textrm{1g}$ phonon. We thereby demonstrate how the rich information from trARPES allows identification of collective modes and their spatial polarization, which explains the mode-dependent coupling to charge order.
\end{abstract}


\maketitle

% ***********************************************************************
Interaction of multiple degrees of freedom in quantum materials can result in ordering phenomena that give rise to complex phase diagrams with competing or intertwined orders \cite{Kivelson_2012}. Such broken-symmetry ground states host characteristic low-energy excitations, which are intimately related to the interactions responsible for ordering. Ultrafast non-equilibrium methods are a new and active field for the investigation of broken-symmetry phases as they grant direct access to collective modes in the time-domain \cite{Tinten_2003, Fritz_2007, Schmitt_2008, Eichenberger_2010, Schaefer_2010, Yusopov_2010} and can provide strong evidence for the ordering mechanism operating in a particular material \cite{Roh11, Hel12}. 

Charge density wave (CDW) formation is one of the most basic and widely studied cases for long-range ordering when only electronic and lattice degrees of freedom are interacting \cite{Gruener_1994}. 
Below a critical temperature $T_\textrm{CDW}$ a spontaneous lattice distortion $\Delta z$ appears, generally originating via the complex interplay of the electronic structure and lattice. This results in a spatial charge modulation at a wave vector $\vec{q}$ (spatial period $2\pi/q$) and opening of a bandgap $2\Delta$ in the band structure, which minimizes the system's total energy by reducing the kinetic energy of the electronic system. One of the characteristic collective excitations in such Peierls-like CDWs is a gapped amplitude mode (AM), which modulates the magnitude $\Delta(t)$ of the order parameter \cite{Gruener_1994}. 

Quasi-2D rare-earth tritellurides $R$Te$_{3}$ ($R$=Y, La-Sm, Gd-Tm) form an incommensurate CDW along the crystallographic $c$-axis and have been studied extensively  \cite{diMasi_1995, Kim_2006, Sacchetti_2006, Ru_2006, Ru_2008, Sacchetti_2009, Tomic_2013, Ban13, Eiter_2013}. 
At low temperatures, the heaviest members of the $R$Te$_3$ series (R=Dy-Tm) additionally order with a second CDW perpendicular to the first CDW \cite{Ru_2008, Moore_2010,Eiter_2013}. Angle-resolved photoemission spectroscopy (ARPES) reveals the Fermi surface (FS) topology and the momentum-dependent gap structure \cite{Gwe98, Brouet_2004, Komoda_2004, Brouet_2008, Moore_2010}.

Collective modes in $R$Te$_{3}$ were studied with Raman spectroscopy \cite{Lav08, Lav10, Eiter_2013}, time-resolved optical reflectivity \cite{Yusopov_2008, Yusopov_2010, Che14}, and time-resolved ARPES (trARPES) \cite{Schmitt_2008, Schmitt_2011, Rettig_2014}. Yet even for this thoroughly investigated CDW system unambiguous identification of an AM is non-trivial. Lowering the translational symmetry in the CDW phase leads to the renormalization of the normal state optical and acoustic phonon branches at $\vec{q}$ \cite{Lav08,Lav10,Yusopov_2008}.
Ubiquitously observed generic phonon modes add to the difficulty of mode identification. Direct comparison of modes observed in the CDW phase to calculated phonon spectra is hampered by the incommensurable CDW in $R$Te$_3$. These challenges are present in CDW materials in general and have led to conflicting reports regarding the AM in $R$Te$_3$ in particular. 
In DyTe$_3$, for instance, a 1.75~THz mode is observed in both temperature-dependent transient reflectivity \cite{Yusopov_2008}, Raman spectroscopy \cite{Lav10} and a trARPES experiment using a pump-pump-probe scheme \cite{Rettig_2014}. Optical measurements attribute this mode to a phonon in the lower symmetry CDW state that is coupled to the AM, whereas trARPES \cite{Rettig_2014} assigns it to a second AM.

In this paper, we report a trARPES study of the collective response of CeTe$_3$ in the CDW phase. The unoccupied band structure is transiently populated by optical transitions, which reveals a gap size of $2\Delta=0.59(2)$~eV. The occupied Te-$5p$ band dispersion near the CDW gap is coherently modified by three modes and analyzed within a transient tight-binding (TB) model. Momentum dependent analysis combined with DFT frozen phonon calculations reveals different coupling to the band structure for in-plane AMs and out-of-plane optical phonons. 

Small rigid shifts of the band dispersion $E_{0}$ are introduced by all three modes, whereas much larger periodic modulations of $\Delta$  are driven by lattice motions $\Omega_{1}=2.2$~THz and $\Omega_{2}=2.7$~THz. We explain this observation by concluding that both modes $\Omega_{1}$ and $\Omega_{2}$ result from in-plane ionic motions which are coupled to the CDW and hence drive oscillations of $\Delta$ as AMs. The $\Omega_{3}=3$~THz mode originates from an out-of-plane optical A$_\textrm{1g}$ phonon that does not couple to the CDW. Our work demonstrates how careful evaluation of the rich information from trARPES allows unambiguous identification of AMs in charge-ordered materials.

\begin{figure}[!h]
\begin{center}
\includegraphics[width=\columnwidth]{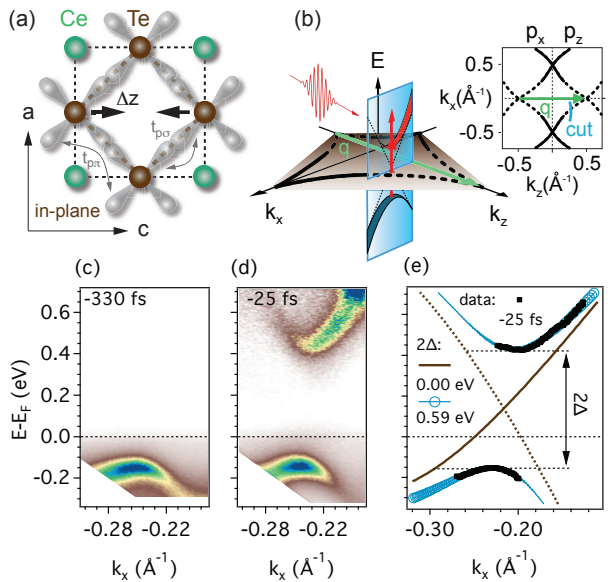}
\caption{
(a) Structure of the Te-planes where the CDW forms along the $c$-axis with lattice displacement $\Delta z$ indicated. The FS is described in a TB model, with the overlap integrals $t_{p\sigma}$ along and $t_{p\pi}$ perpendicular to Te-$5p$ orbitals. 
(b) Cartoon of the trARPES experiment on CeTe$_{3}$ in the CDW region of the FS (\textit{dotted lines}). The \textit{blue plane} illustrates an energy versus momentum cut with Te-$5p$ bands and a pump pulse exciting electrons into unoccupied states. The \textit{blue line} indicates the momentum cut in the experiment across the gapped region (\textit{dotted lines}) of the calculated 2D FS, \textit{green arrow} denotes the CDW wave vector $\vec{q}$. trARPES spectra for selected pump-probe delays, 330 fs before the arrival of the pump pulse (c) and during the presence of the pump pulse (d). Intensities are rescaled exponentially as function of energy for enhanced visibility of the transiently occupied features above $E_\textrm{F}$. 
(e) TB fit (\textit{blue markers}) according to Eqs.~(\ref{Bragg_1}) and (\ref{TB_1}) to the extracted band dispersion (\textit{black markers}) at $-25$~fs yields $2\Delta = 0.59(2)$~eV. Size of the $blue$ $markers$ is proportional to the calculated spectral weight of the TB bands \cite{Brouet_2008}. \textit{Brown lines} indicate the calculated metallic bare band dispersion.
}\label{Fig1}
\end{center}
\end{figure}

Our trARPES setup \cite{Sobota_2012,Yang_2014} consists of an amplified laser system operating at $310$~kHz repetition rate that drives an optical parametric amplifier to optically excite the sample with $\sim50$~fs pump pulses at a photon energy of $1.03$~eV. Electrons are photoemitted with $\sim150$~fs probe pulses at $6$~eV as function of pump-probe delay and collected with 50~fs delay steps in a hemispherical electron analyzer, yielding a total energy resolution of $\sim22$~meV.  CeTe$_{3}$ single crystals were grown by slow cooling of a binary melt \cite{Ru_2006}. Measurements were performed on freshly cleaved CeTe$_{3}$ samples at $T=100$~K in an ultrahigh vacuum chamber with a pressure of $<1\times 10^{-10}$~torr. Phonon eigenmodes of the normal state of $R$Te$_3$ were calculated from the dynamical matrix using the Quantum-ESPRESSO \cite{Giannozzi_2009} package with ultrasoft pseudopotentials and the Perdew-Burke-Ernzerhof \cite{Perdew_1996} exchange-correlation functional, a kinetic energy cutoff of $500$~eV, and 13$\times$3$\times$13 $k$-points. Incident pump fluences of $55-354$~$\mu$J/cm$^{2}$ correspond to absorbed energies of $16-103$~meV per unit cell at the surface \cite{sup_mat}. 2$\Delta$ decreases by no more than $15\%$ and the system responds linearly \cite{sup_mat} without melting of the charge order. This is in contrast to previous studies at higher excitation densities \cite{Schmitt_2008, Schmitt_2011, Rettig_2014}, where qualitative changes in the dispersion prevent decoupling the effects of coherent in-plane and out-of-plane motions. 

Fig.\ref{Fig1}(b) illustrates the cut through the FS where the trARPES data, which are shown in Fig. \ref{Fig1}(c) and (d) for selected pump-probe delays, have been taken. Due to an asymmetric band dispersion and a slightly curved cut in the $\vec{k}=\{ k_x, k_z \}$ plane the Fermi momentum $k_\textrm{F}$ does not exactly coincide with the maximum (minimum) position of the occupied (unoccupied) band. For simplicity, dispersions along $\vec{k}$ are projected on $k_x$. At a pump-probe delay of $-330$~fs the Te-$5p$ band with the occupied part of the CDW gap is observed in equilibrium. At $-25$~fs, the leading edge of the pump pulse promotes electrons into unoccupied Te-$5p$ states without yet changing the band dispersion significantly, which occurs at later delays. The magnitude of the full CDW gap  $2\Delta=0.59(2)$~eV is directly given by the difference of the maximum of the lower band at $E-E_\textrm{F}=-0.16(1)$~eV and the minimum of the upper band at $E-E_\textrm{F} = 0.43(1)$~eV. This is in agreement with previous results from optical reflectivity \cite{Sacchetti_2006, Sacchetti_2009} while being smaller than extracted from ARPES \cite{Brouet_2004} and larger than estimated by tunneling experiments \cite{Tomic_2013}. 

\begin{figure}
\begin{center}
\includegraphics[width=\columnwidth]{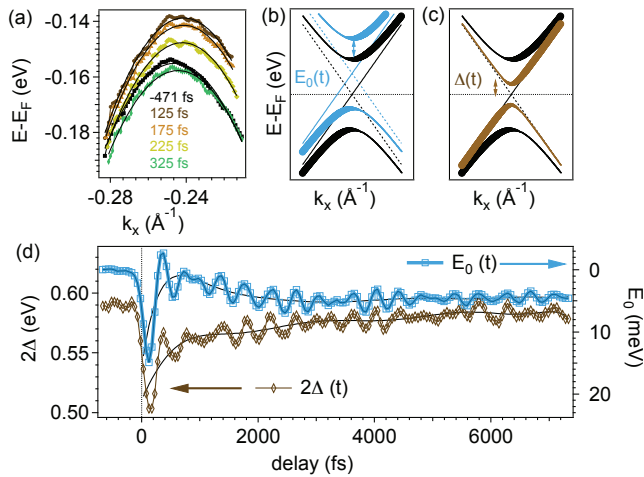}
\caption{
(a) Measured dispersion of the occupied Te-$5p$ band for selected delays (\textit{markers}) and transient TB fits (\textit{lines}). 
Sketched effect of a transient rigid energy shift $E_0(t)$ (b) and change of the CDW gap $\Delta(t)$ (c) on the band dispersion.
(d) 2$\Delta(t)$ and $E_0(t)$ with smooth backgrounds (\textit{black lines}) of a 10$^{th}$ order polynomial indicated.
}\label{Fig2}
\end{center}
\end{figure}

To obtain a quantitative description of the electronic bands near $E_\textrm{F}$ we consider a TB model of the in-plane Te-$5p_{x/z}$ orbitals \cite{Brouet_2008, Eiter_2013} as sketched in Fig.~\ref{Fig1}(a). Nesting of the metallic bare bands $\epsilon(\vec{k})$ via  $\vec{q}$ creates two branches $E_{1,2}(\vec{k})$, which are separated by $2\Delta$ \cite{Gruener_1994}:

\begin{align}
\label{Bragg_1}
E_{1,2}(\vec{k})=\frac{\epsilon_{\vec{k}}+\epsilon_{\vec{k}-\vec{q}}}{2} 
                 \pm\sqrt{\left(\frac{\epsilon_{\vec{k}}-\epsilon_{\vec{k}-\vec{q}}}{2} \right)^{2}+\Delta^{2}}. 
\end{align}
$\epsilon(\vec{k})$ is approximated by a non-interacting TB model \cite{Brouet_2008, Eiter_2013} of the in-plane Te-$5p$ orbitals, see Fig.~\ref{Fig1}(a).
\begin{align}
\label{TB_1}
\epsilon(\vec{k})= 	& -2 t_{p\sigma} \cos((k_{x}-k_{z})a/2) \\
					& -2 t_{p\pi}\cos((k_{x}+k_{z})a/2) - 
                                   E_\textrm{F} + E_0  \notag \quad. 
\end{align}
The TB coupling parameters $t_{p\sigma} =-1.9$~eV along and $t_{p\pi}=0.35$~eV perpendicular to the Te-chains (Fig.~\ref{Fig1}(a)) have been established by  ARPES experiments \cite{Brouet_2004, Brouet_2008, Moore_2010}. $E_\textrm{F} = -2 t_{p\sigma}\sin{(\pi/8)}$ and $\left|\vec{q}\right|=0.685\,(2\pi/a)$ are constant. The in-plane lattice spacing is given by $a=4.34$~\AA~\cite{Ru_2008}. The band dispersion is extracted by fitting a Gauss function to the spectra as function of $\vec{k}$, and yield the data points in Fig.~\ref{Fig1}(e). Eqs. (\ref{Bragg_1}) and (\ref{TB_1}) are combined and simultaneously fitted to the extracted dispersion above and below $E_\textrm{F}$ at $-25$~fs as shown in Fig.~\ref{Fig1}(e). The rigid energy shift $E_0=0$ is kept fixed and the only free fit parameter is $\Delta$. This yields $2\Delta=0.59(1)$~eV, which is identical to the value derived from the difference of upper and lower band energies at $-25$~fs. 

After having established that the TB model is well suited to describe the gapped dispersion near $E_\textrm{F}$, it is applied to the occupied band for all delays, as shown in Fig.~\ref{Fig2}(a), with $\Delta(t)$ and $E_0(t)$ now being free time-dependent fit parameters. The low excitation regime allows for holding the TB parameters $t_{p\sigma}$ and $t_{p\pi}$ constant. $\Delta(t)$ leads to momentum-dependent modulations in the spectra, and is naturally sensitive to the AM \cite{Gruener_1994} as illustrated in Fig.~\ref{Fig2}(c). In contrast, $E_0(t)$ is a momentum-independent energy shift of the whole band structure, as sketched in Fig.~\ref{Fig2}(b). As we will show, these parameters capture the two types of modes observed.

Results for $2\Delta$(t) and $E_0(t)$ are shown in Fig.~\ref{Fig2}(d). The magnitude of $E_0(t)$ amounts to a few meV and is much smaller than the changes in $2\Delta(t)$, which are on the order of several $10$~meV. Within the pump pulse duration, $2\Delta$ drops by $15$~$\%$ yet the charge order is only perturbed and not destroyed \cite{sup_mat}. After $\sim 3$~ps, the system has thermalized at reduced gap size, which agrees with the timescale of the suppression of the structural order parameter satellite observed in time-resolved electron diffraction \cite{Han12, Tao13}.

\begin{figure}[h!]
\begin{center}
\includegraphics[width=\columnwidth]{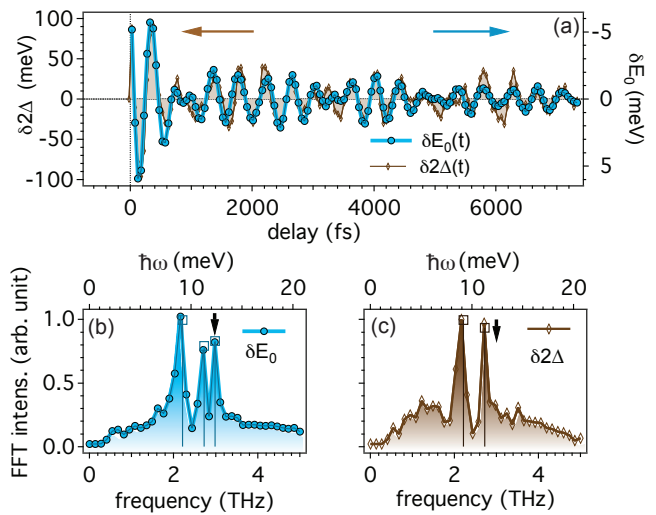}
\caption{
(a) Residuals of the polynomial background fit shown in Fig.~\ref{Fig2}(d). (b) and (c) FT of (a). The $vertical$ $lines$ and $markers$ at $\Omega_1 = 2.19(1)$~THz, $\Omega_2 = 2.69(1)$~THz and $\Omega_3 = 2.98(1)$~THz indicate mode frequencies and amplitudes independently obtained from fitting damped cosine functions \cite{sup_mat}.
}\label{Fig3}
\end{center}
\end{figure}

The pronounced beating patterns of $2\Delta(t)$ and $E_0(t)$ exhibit an oscillatory response with similar yet distinct frequencies. For a more detailed analysis smooth backgrounds ($10^{th}$ order polynomials) are subtracted and the residuals $\delta2\Delta(t)$ and $\delta E_0(t)$ are plotted in Fig.~\ref{Fig3}(a). Analysis via Fourier transformation (FT) of $\delta2\Delta(t)$ reveals two dominant frequencies at $\Omega_1 = 2.2(1)$~THz and $\Omega_2 = 2.7(1)$~THz, as shown in Fig.~\ref{Fig3}(c). In contrast, the oscillatory response and particularly the beating pattern of $\delta E_0(t)$ differs from $\delta2\Delta$(t). In addition to $\Omega_1$ and $\Omega_2$, the FT of $\delta E_0(t)$ exhibits a third pronounced peak at $\Omega_3 = 3.0(1)$~THz, marked by a vertical arrow in Fig. \ref{Fig3}(b). These findings are robust regardless of the background subtraction \cite{sup_mat}. Fitting damped cosine functions to $\delta2\Delta(t)$ and $\delta E_0(t)$ \cite{sup_mat} yields the same frequencies albeit with higher resolution as indicated in Fig.~\ref{Fig3}. 

\begin{figure}
\begin{center}
\includegraphics[width = \columnwidth]{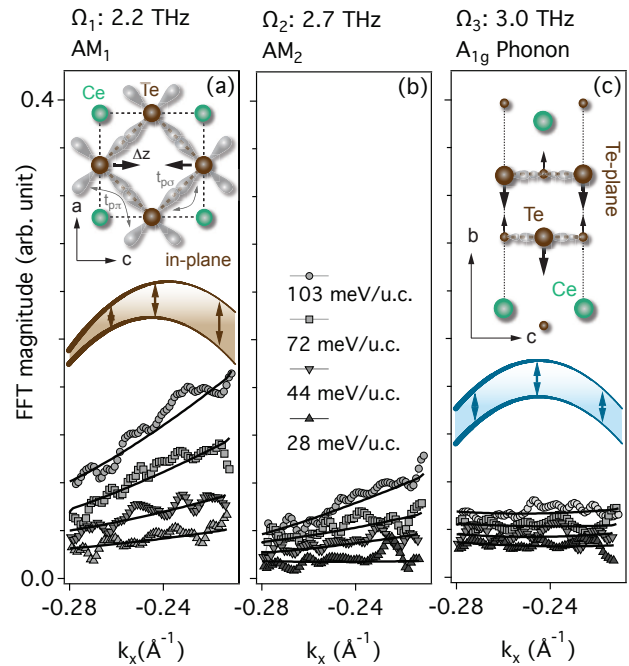}
\caption{
Mode-dependent modulations of the Te-$5p$ bands ($markers$) by the three dominant modes is derived from momentum-dependent FT amplitudes for deposited energies of $28-103$~meV per unit cell. $Solid$ $lines$ are obtained from the corresponding TB fits of the transient Te-$5p$ bands \cite{sup_mat}.  
(a) The AM at $\Omega_1=2.2$~THz modulates the in-plane lattice displacement $\Delta z$ and leads to a strongly momentum- and fluence-dependent response. 
(b) The AM at $\Omega_2 = 2.7(1)$~THz exhibits a weaker momentum- and fluence-dependence, which indicates a weaker influence on the in-plane CDW.
(c) The out-of-plane A$_{1g}$ mode at $\Omega_3 = 3$~THz only shifts the band rigidly as it does not couple to the in-plane charge order.
}\label{Fig4}
\end{center}
\end{figure}

The mode at $\Omega_1$ is identified as a previously observed AM of $R$Te$_3$. As sketched in Fig.~\ref{Fig4}(a), coherent in-plane atomic motions modulate the lattice distortion $\Delta z$ and thus directly affect $\Delta$. This assignment is supported by comparing frequency $\Omega_1$ to the soft mode observed in temperature-dependent transient optical reflectivity \cite{Yusopov_2008,Yusopov_2010}, temperature-dependent Raman spectroscopy \cite{Lav08, Lav10, Eiter_2013}, and temperature- and fluence-dependent trARPES \cite{Schmitt_2008, Schmitt_2011, Rettig_2014}. 

The mode at $\Omega_{2}$ is missing at the zone center of the frozen phonon calculations, suggesting it originates either from a normal state optical or acoustic branch which is renormalized by coupling to the charge modulation at $\vec{q}$. A finite projection of the ionic motion on the lattice displacement $\Delta z$ associated with the CDW then naturally explains its presence in $\Delta(t)$ and the temperature-dependent softening in \cite{Lav10,Yusopov_2008}. We thus assign $\Omega_{2}$ as a 2nd AM.

The mode at $\Omega_3$ is assigned to an out-of-plane A$_{1g}$ phonon, which is not related to the CDW by comparison to frozen phonon calculations and transient reflectivity \cite{Yusopov_2008}. The calculated displacement of the Te-atoms is sketched in Fig.~\ref{Fig4}(c). The response of the Te-$5p$ band is remarkably mode-dependent; $\Omega_3$ mainly affects $E_0$, and has a much smaller influence on $\Delta$ even though $\Omega_3$ involves the motion of the Te atoms, which form the CDW bands. 
However, the out-of-plane motion couples only weakly to the charge order, with correspondingly small influence on the order parameter $\Delta$.

To further investigate this explanation for mode-dependent coupling we plot FT amplitudes for all three modes in Fig.~\ref{Fig4} as function of fluence and momentum. These are derived from an FT analysis of the oscillatory binding energies, which have been extracted from transient spectra as function of $\vec{k}$ \cite{sup_mat}. The influence on the dispersion is illustrated in Fig.~\ref{Fig4}(a) and (c). The AM at $\Omega_1$ is strongly fluence- and momentum-dependent as expected for a reduction of $\Delta$. The AM at $\Omega_2$ exhibits a qualitatively similar behavior albeit of smaller magnitude. 
This is consistent with a smaller coupling between the phonon with the charge density or a smaller projection of the ionic motion on the lattice distortion $\Delta z$. 

In contrast, the out-of-plane A$_{1g}$ phonon at $\Omega_3$ causes a strikingly $k$-independent shift, which is only weakly dependent on fluence. This suggests a small fluence dependent atomic displacement $\mu(t)$ and a constant deformation potential $D_{5p}(\vec{k})$ that explains the rigid energy shift via $E_0(t) \sim D_{5p}(\vec{k}) \mu(t)$ \cite{Khan_1984, Fauvre_2013}. We confirm the approximation of a k-independent deformation potential by DFT calculations for atomic displacements $\mu$ of up to 0.13~\AA, shown in Ref. \cite{sup_mat}.

In summary, trARPES measurements on CeTe$_{3}$ reveal a CDW gap size of $2\Delta=0.59$~eV. Coherent excitations of two AMs at $\Omega_{1}=2.2$~THz and at $\Omega_2=2.7$~THz and a generic A$_{1g}$ phonon at $\Omega_3=3$~THz are identified by fits of a time-dependent model dispersion and comparison to normal state frozen phonon calculations. The out-of-plane A$_{1g}$ phonon leads to small rigid shifts of the entire Te-$5p$ band but does not affect $\Delta$. In contrast, $\Omega_1$ and $\Omega_2$ result in pronounced changes of $\Delta$. We point out that our $k$-dependent analysis combined with low excitation densities is key to identifying and assigning AMs. We have generalized our approach and introduced a model-free evaluation (Fig~\ref{Fig4}), which can be applied to materials where a detailed description of the band dispersion may not be available. More specifically, our approach can be transfered to charge-ordering in striped cuprates \cite{Tranquada_1995}, which has identical symmetry to $R$Te$_3$ \cite{Yao_2006}. 

\begin{acknowledgments}
We acknowledge helpful discussions with B. Moritz, S. Gerber and R. Hackl. This work was supported by the U.S. Department of Energy, Office of Science, Basic Energy Sciences, Materials Sciences and Engineering Division under contract DE-AC02-76SF00515. D.L. acknowledges support from the Swiss National Science Foundation, under the Fellowship number P300P2$\textunderscore$151328. S.-L.Y. acknowledges support from the Stanford Graduate Fellowship. The work by A.F.K. was supported by the Laboratory Directed Research and Development Program of Lawrence Berkeley National Laboratory under the U.S. Department of Energy contract number DE-AC02-05CH11231.  
\end{acknowledgments}

% ***** BIBLIOGRAPHY


\begin{thebibliography}{41}
\expandafter\ifx\csname natexlab\endcsname\relax\def\natexlab#1{#1}\fi
\expandafter\ifx\csname bibnamefont\endcsname\relax
  \def\bibnamefont#1{#1}\fi
\expandafter\ifx\csname bibfnamefont\endcsname\relax
  \def\bibfnamefont#1{#1}\fi
\expandafter\ifx\csname citenamefont\endcsname\relax
  \def\citenamefont#1{#1}\fi
\expandafter\ifx\csname url\endcsname\relax
  \def\url#1{\texttt{#1}}\fi
\expandafter\ifx\csname urlprefix\endcsname\relax\def\urlprefix{URL }\fi
\providecommand{\bibinfo}[2]{#2}
\providecommand{\eprint}[2][]{\url{#2}}

\bibitem[{\citenamefont{Fradkin and Kivelson}(2012)}]{Kivelson_2012}
\bibinfo{author}{\bibfnamefont{E.}~\bibnamefont{Fradkin}} \bibnamefont{and}
  \bibinfo{author}{\bibfnamefont{S.~A.} \bibnamefont{Kivelson}},
  \bibinfo{journal}{Nature Physics} \textbf{\bibinfo{volume}{8}},
  \bibinfo{pages}{864} (\bibinfo{year}{2012}).

\bibitem[{\citenamefont{Sokolowski-Tinten
  et~al.}(2003)\citenamefont{Sokolowski-Tinten, Blome, Blums, Cavalleri,
  Dietrich, Tarasevitch, Uschmann, F\"orster, Kammler, von Hoegen
  et~al.}}]{Tinten_2003}
\bibinfo{author}{\bibfnamefont{K.}~\bibnamefont{Sokolowski-Tinten}},
  \bibinfo{author}{\bibfnamefont{C.}~\bibnamefont{Blome}},
  \bibinfo{author}{\bibfnamefont{J.}~\bibnamefont{Blums}},
  \bibinfo{author}{\bibfnamefont{A.}~\bibnamefont{Cavalleri}},
  \bibinfo{author}{\bibfnamefont{C.}~\bibnamefont{Dietrich}},
  \bibinfo{author}{\bibfnamefont{A.}~\bibnamefont{Tarasevitch}},
  \bibinfo{author}{\bibfnamefont{I.}~\bibnamefont{Uschmann}},
  \bibinfo{author}{\bibfnamefont{E.}~\bibnamefont{F\"orster}},
  \bibinfo{author}{\bibfnamefont{M.}~\bibnamefont{Kammler}},
  \bibinfo{author}{\bibfnamefont{M.~H.} \bibnamefont{von Hoegen}},
  \bibnamefont{et~al.}, \bibinfo{journal}{Nature}
  \textbf{\bibinfo{volume}{422}}, \bibinfo{pages}{287} (\bibinfo{year}{2003}).

\bibitem[{\citenamefont{Fritz et~al.}(2007)\citenamefont{Fritz, Reis, Adams,
  Akre, Arthur, Blome, Bucksbaum, Cavalieri, Engemann, Fahy
  et~al.}}]{Fritz_2007}
\bibinfo{author}{\bibfnamefont{D.~M.} \bibnamefont{Fritz}},
  \bibinfo{author}{\bibfnamefont{D.~A.} \bibnamefont{Reis}},
  \bibinfo{author}{\bibfnamefont{B.}~\bibnamefont{Adams}},
  \bibinfo{author}{\bibfnamefont{R.~A.} \bibnamefont{Akre}},
  \bibinfo{author}{\bibfnamefont{J.}~\bibnamefont{Arthur}},
  \bibinfo{author}{\bibfnamefont{C.}~\bibnamefont{Blome}},
  \bibinfo{author}{\bibfnamefont{P.~H.} \bibnamefont{Bucksbaum}},
  \bibinfo{author}{\bibfnamefont{A.~L.} \bibnamefont{Cavalieri}},
  \bibinfo{author}{\bibfnamefont{S.}~\bibnamefont{Engemann}},
  \bibinfo{author}{\bibfnamefont{S.}~\bibnamefont{Fahy}}, \bibnamefont{et~al.},
  \bibinfo{journal}{Science} \textbf{\bibinfo{volume}{315}},
  \bibinfo{pages}{633} (\bibinfo{year}{2007}).

\bibitem[{\citenamefont{Schmitt et~al.}(2008)\citenamefont{Schmitt, Kirchmann,
  Bovensiepen, Moore, Rettig, Krenz, Chu, Ru, Perfetti, Lu
  et~al.}}]{Schmitt_2008}
\bibinfo{author}{\bibfnamefont{F.}~\bibnamefont{Schmitt}},
  \bibinfo{author}{\bibfnamefont{P.~S.} \bibnamefont{Kirchmann}},
  \bibinfo{author}{\bibfnamefont{U.}~\bibnamefont{Bovensiepen}},
  \bibinfo{author}{\bibfnamefont{R.~G.} \bibnamefont{Moore}},
  \bibinfo{author}{\bibfnamefont{L.}~\bibnamefont{Rettig}},
  \bibinfo{author}{\bibfnamefont{M.}~\bibnamefont{Krenz}},
  \bibinfo{author}{\bibfnamefont{J.-H.} \bibnamefont{Chu}},
  \bibinfo{author}{\bibfnamefont{N.}~\bibnamefont{Ru}},
  \bibinfo{author}{\bibfnamefont{L.}~\bibnamefont{Perfetti}},
  \bibinfo{author}{\bibfnamefont{D.~H.} \bibnamefont{Lu}},
  \bibnamefont{et~al.}, \bibinfo{journal}{Science}
  \textbf{\bibinfo{volume}{321}}, \bibinfo{pages}{1649} (\bibinfo{year}{2008}).

\bibitem[{\citenamefont{Eichberger et~al.}(2010)\citenamefont{Eichberger,
  Sch\"afer, Krumova, Beyer, Demsar, Berger, Moriena, Sciaini, and
  Miller}}]{Eichenberger_2010}
\bibinfo{author}{\bibfnamefont{M.}~\bibnamefont{Eichberger}},
  \bibinfo{author}{\bibfnamefont{H.}~\bibnamefont{Sch\"afer}},
  \bibinfo{author}{\bibfnamefont{M.}~\bibnamefont{Krumova}},
  \bibinfo{author}{\bibfnamefont{M.}~\bibnamefont{Beyer}},
  \bibinfo{author}{\bibfnamefont{J.}~\bibnamefont{Demsar}},
  \bibinfo{author}{\bibfnamefont{H.}~\bibnamefont{Berger}},
  \bibinfo{author}{\bibfnamefont{G.}~\bibnamefont{Moriena}},
  \bibinfo{author}{\bibfnamefont{G.}~\bibnamefont{Sciaini}}, \bibnamefont{and}
  \bibinfo{author}{\bibfnamefont{R.~J.~D.}~\bibnamefont{Miller}},
  \bibinfo{journal}{Nature} \textbf{\bibinfo{volume}{468}},
  \bibinfo{pages}{799} (\bibinfo{year}{2010}).

\bibitem[{\citenamefont{Sch\"afer et~al.}(2010)\citenamefont{Sch\"afer,
  Kabanov, Beyer, Biljakovic, and Demsar}}]{Schaefer_2010}
\bibinfo{author}{\bibfnamefont{H.}~\bibnamefont{Sch\"afer}},
  \bibinfo{author}{\bibfnamefont{V.~V.} \bibnamefont{Kabanov}},
  \bibinfo{author}{\bibfnamefont{M.}~\bibnamefont{Beyer}},
  \bibinfo{author}{\bibfnamefont{K.}~\bibnamefont{Biljakovic}},
  \bibnamefont{and} \bibinfo{author}{\bibfnamefont{J.}~\bibnamefont{Demsar}},
  \bibinfo{journal}{Phys. Rev. Lett.} \textbf{\bibinfo{volume}{105}},
  \bibinfo{pages}{066402} (\bibinfo{year}{2010}).

\bibitem[{\citenamefont{Yusupov et~al.}(2010)\citenamefont{Yusupov, Mertelj,
  Kabanov, Brazovskii, Kusar, Chu, Fisher, and Mihailovic}}]{Yusopov_2010}
\bibinfo{author}{\bibfnamefont{R.}~\bibnamefont{Yusupov}},
  \bibinfo{author}{\bibfnamefont{T.}~\bibnamefont{Mertelj}},
  \bibinfo{author}{\bibfnamefont{V.~V.} \bibnamefont{Kabanov}},
  \bibinfo{author}{\bibfnamefont{S.}~\bibnamefont{Brazovskii}},
  \bibinfo{author}{\bibfnamefont{P.}~\bibnamefont{Kusar}},
  \bibinfo{author}{\bibfnamefont{J.-H.} \bibnamefont{Chu}},
  \bibinfo{author}{\bibfnamefont{I.~R.} \bibnamefont{Fisher}},
  \bibnamefont{and}
  \bibinfo{author}{\bibfnamefont{D.}~\bibnamefont{Mihailovic}},
  \bibinfo{journal}{Nature Physics} \textbf{\bibinfo{volume}{6}},
  \bibinfo{pages}{681} (\bibinfo{year}{2010}).

\bibitem[{\citenamefont{Rohwer et~al.}(2011)\citenamefont{Rohwer, Hellmann,
  Wiesenmayer, Sohrt, Stange, Slomski, Carr, Liu, Avila, Kall\"ane
  et~al.}}]{Roh11}
\bibinfo{author}{\bibfnamefont{T.}~\bibnamefont{Rohwer}},
  \bibinfo{author}{\bibfnamefont{S.}~\bibnamefont{Hellmann}},
  \bibinfo{author}{\bibfnamefont{M.}~\bibnamefont{Wiesenmayer}},
  \bibinfo{author}{\bibfnamefont{C.}~\bibnamefont{Sohrt}},
  \bibinfo{author}{\bibfnamefont{A.}~\bibnamefont{Stange}},
  \bibinfo{author}{\bibfnamefont{B.}~\bibnamefont{Slomski}},
  \bibinfo{author}{\bibfnamefont{A.}~\bibnamefont{Carr}},
  \bibinfo{author}{\bibfnamefont{Y.}~\bibnamefont{Liu}},
  \bibinfo{author}{\bibfnamefont{L.~M.} \bibnamefont{Avila}},
  \bibinfo{author}{\bibfnamefont{M.}~\bibnamefont{Kall\"ane}},
  \bibnamefont{et~al.}, \bibinfo{journal}{Nature}
  \textbf{\bibinfo{volume}{471}}, \bibinfo{pages}{490} (\bibinfo{year}{2011}).

\bibitem[{\citenamefont{Hellmann et~al.}(2012)\citenamefont{Hellmann, Rohwer,
  Kall\"ne, Hanff, Sohrt, Stange, Carr, Murnane, Kapteyn, Kipp
  et~al.}}]{Hel12}
\bibinfo{author}{\bibfnamefont{S.}~\bibnamefont{Hellmann}},
  \bibinfo{author}{\bibfnamefont{T.}~\bibnamefont{Rohwer}},
  \bibinfo{author}{\bibfnamefont{M.}~\bibnamefont{Kall\"ane}},
  \bibinfo{author}{\bibfnamefont{K.}~\bibnamefont{Hanff}},
  \bibinfo{author}{\bibfnamefont{C.}~\bibnamefont{Sohrt}},
  \bibinfo{author}{\bibfnamefont{A.}~\bibnamefont{Stange}},
  \bibinfo{author}{\bibfnamefont{A.}~\bibnamefont{Carr}},
  \bibinfo{author}{\bibfnamefont{M.~M.} \bibnamefont{Murnane}},
  \bibinfo{author}{\bibfnamefont{H.~C.} \bibnamefont{Kapteyn}},
  \bibinfo{author}{\bibfnamefont{L.}~\bibnamefont{Kipp}}, \bibnamefont{et~al.},
  \bibinfo{journal}{Nature Commun.} \textbf{\bibinfo{volume}{3}},
  \bibinfo{pages}{1069} (\bibinfo{year}{2012}).

\bibitem[{\citenamefont{Gr\"uner}(1994)}]{Gruener_1994}
\bibinfo{author}{\bibfnamefont{G.}~\bibnamefont{Gr\"uner}},
  \emph{\bibinfo{title}{Density Waves in Solids}}, vol.~\bibinfo{volume}{89}
  (\bibinfo{publisher}{Addison-Wesley}, \bibinfo{year}{1994}).

\bibitem[{\citenamefont{DiMasi et~al.}(1995)\citenamefont{DiMasi, Aronson,
  Mansfield, Foran, and Lee}}]{diMasi_1995}
\bibinfo{author}{\bibfnamefont{E.~D.} \bibnamefont{DiMasi}},
  \bibinfo{author}{\bibfnamefont{M.~C.} \bibnamefont{Aronson}},
  \bibinfo{author}{\bibfnamefont{J.~F.} \bibnamefont{Mansfield}},
  \bibinfo{author}{\bibfnamefont{B.}~\bibnamefont{Foran}}, \bibnamefont{and}
  \bibinfo{author}{\bibfnamefont{S.}~\bibnamefont{Lee}},
  \bibinfo{journal}{Phys. Rev. B} \textbf{\bibinfo{volume}{52}},
  \bibinfo{pages}{14516} (\bibinfo{year}{1995}).

\bibitem[{\citenamefont{Kim et~al.}(2006)\citenamefont{Kim, Malliakas, Tomic,
  Tessmer, Kanatzidis, and Billinge}}]{Kim_2006}
\bibinfo{author}{\bibfnamefont{H.~J.} \bibnamefont{Kim}},
  \bibinfo{author}{\bibfnamefont{C.~D.} \bibnamefont{Malliakas}},
  \bibinfo{author}{\bibfnamefont{A.~T.} \bibnamefont{Tomic}},
  \bibinfo{author}{\bibfnamefont{S.~H.} \bibnamefont{Tessmer}},
  \bibinfo{author}{\bibfnamefont{M.~G.} \bibnamefont{Kanatzidis}},
  \bibnamefont{and} \bibinfo{author}{\bibfnamefont{S.~J.~L.}
  \bibnamefont{Billinge}}, \bibinfo{journal}{Phys. Rev. Lett.}
  \textbf{\bibinfo{volume}{96}}, \bibinfo{pages}{226401}
  (\bibinfo{year}{2006}).

\bibitem[{\citenamefont{Sacchetti et~al.}(2006)\citenamefont{Sacchetti,
  Degiorgi, Giamarchi, Ru, and Fisher}}]{Sacchetti_2006}
\bibinfo{author}{\bibfnamefont{A.}~\bibnamefont{Sacchetti}},
  \bibinfo{author}{\bibfnamefont{L.}~\bibnamefont{Degiorgi}},
  \bibinfo{author}{\bibfnamefont{T.}~\bibnamefont{Giamarchi}},
  \bibinfo{author}{\bibfnamefont{N.}~\bibnamefont{Ru}}, \bibnamefont{and}
  \bibinfo{author}{\bibfnamefont{I.~R.} \bibnamefont{Fisher}},
  \bibinfo{journal}{Phys. Rev. B} \textbf{\bibinfo{volume}{74}},
  \bibinfo{pages}{125115} (\bibinfo{year}{2006}).

\bibitem[{\citenamefont{Ru and Fisher}(2006)}]{Ru_2006}
\bibinfo{author}{\bibfnamefont{N.}~\bibnamefont{Ru}} \bibnamefont{and}
  \bibinfo{author}{\bibfnamefont{I.~R.} \bibnamefont{Fisher}},
  \bibinfo{journal}{Phys. Rev. B} \textbf{\bibinfo{volume}{73}},
  \bibinfo{pages}{033101} (\bibinfo{year}{2006}).

\bibitem[{\citenamefont{Ru et~al.}(2008)\citenamefont{Ru, Condron, Margulis,
  Shin, Laverock, Dugdale, Toney, and Fisher}}]{Ru_2008}
\bibinfo{author}{\bibfnamefont{N.}~\bibnamefont{Ru}},
  \bibinfo{author}{\bibfnamefont{C.~L.} \bibnamefont{Condron}},
  \bibinfo{author}{\bibfnamefont{G.~Y.} \bibnamefont{Margulis}},
  \bibinfo{author}{\bibfnamefont{K.~Y.} \bibnamefont{Shin}},
  \bibinfo{author}{\bibfnamefont{J.}~\bibnamefont{Laverock}},
  \bibinfo{author}{\bibfnamefont{S.~B.} \bibnamefont{Dugdale}},
  \bibinfo{author}{\bibfnamefont{M.~F.} \bibnamefont{Toney}}, \bibnamefont{and}
  \bibinfo{author}{\bibfnamefont{I.~R.} \bibnamefont{Fisher}},
  \bibinfo{journal}{Phys. Rev. B} \textbf{\bibinfo{volume}{77}},
  \bibinfo{pages}{035114} (\bibinfo{year}{2008}).

\bibitem[{\citenamefont{Sacchetti et~al.}(2009)\citenamefont{Sacchetti,
  Condron, Gvasaliya, Pfuner, Lavagnini, Baldini, Toney, Merlini, Hanfland,
  Mesot et~al.}}]{Sacchetti_2009}
\bibinfo{author}{\bibfnamefont{A.}~\bibnamefont{Sacchetti}},
  \bibinfo{author}{\bibfnamefont{C.~L.} \bibnamefont{Condron}},
  \bibinfo{author}{\bibfnamefont{S.~N.} \bibnamefont{Gvasaliya}},
  \bibinfo{author}{\bibfnamefont{F.}~\bibnamefont{Pfuner}},
  \bibinfo{author}{\bibfnamefont{M.}~\bibnamefont{Lavagnini}},
  \bibinfo{author}{\bibfnamefont{M.}~\bibnamefont{Baldini}},
  \bibinfo{author}{\bibfnamefont{M.~F.} \bibnamefont{Toney}},
  \bibinfo{author}{\bibfnamefont{M.}~\bibnamefont{Merlini}},
  \bibinfo{author}{\bibfnamefont{M.}~\bibnamefont{Hanfland}},
  \bibinfo{author}{\bibfnamefont{J.}~\bibnamefont{Mesot}},
  \bibnamefont{et~al.}, \bibinfo{journal}{Phys. Rev. B}
  \textbf{\bibinfo{volume}{79}}, \bibinfo{pages}{201101(R)}
  (\bibinfo{year}{2009}).

\bibitem[{\citenamefont{Tomic et~al.}(2009)\citenamefont{Tomic, Rak, Veazey,
  Malliakas, Mahanti, Kanatzidis, and Tessmer}}]{Tomic_2013}
\bibinfo{author}{\bibfnamefont{A.}~\bibnamefont{Tomic}},
  \bibinfo{author}{\bibfnamefont{Z.}~\bibnamefont{Rak}},
  \bibinfo{author}{\bibfnamefont{J.~P.} \bibnamefont{Veazey}},
  \bibinfo{author}{\bibfnamefont{C.~D.} \bibnamefont{Malliakas}},
  \bibinfo{author}{\bibfnamefont{S.~D.} \bibnamefont{Mahanti}},
  \bibinfo{author}{\bibfnamefont{M.~G.} \bibnamefont{Kanatzidis}},
  \bibnamefont{and} \bibinfo{author}{\bibfnamefont{S.~H.}
  \bibnamefont{Tessmer}}, \bibinfo{journal}{Phys. Rev. B}
  \textbf{\bibinfo{volume}{79}}, \bibinfo{pages}{085422}
  (\bibinfo{year}{2009}).

\bibitem[{\citenamefont{Banerjee et~al.}(2013)\citenamefont{Banerjee, Feng,
  Silevitch, Wang, Lang, Kuo, Fisher, and Rosenbaum}}]{Ban13}
\bibinfo{author}{\bibfnamefont{A.}~\bibnamefont{Banerjee}},
  \bibinfo{author}{\bibfnamefont{Y.}~\bibnamefont{Feng}},
  \bibinfo{author}{\bibfnamefont{D.~M.} \bibnamefont{Silevitch}},
  \bibinfo{author}{\bibfnamefont{J.}~\bibnamefont{Wang}},
  \bibinfo{author}{\bibfnamefont{J.~C.} \bibnamefont{Lang}},
  \bibinfo{author}{\bibfnamefont{H.-H.} \bibnamefont{Kuo}},
  \bibinfo{author}{\bibfnamefont{I.~R.} \bibnamefont{Fisher}},
  \bibnamefont{and} \bibinfo{author}{\bibfnamefont{T.~F.}
  \bibnamefont{Rosenbaum}}, \bibinfo{journal}{Phys. Rev. B}
  \textbf{\bibinfo{volume}{87}}, \bibinfo{pages}{155131}
  (\bibinfo{year}{2013}).

\bibitem[{\citenamefont{Eiter et~al.}(2013)\citenamefont{Eiter, Lavagnini,
  Hackl, Nowadnick, Kemper, Devereaux, Chu, Analytis, Fisher, and
  Degiorgi}}]{Eiter_2013}
\bibinfo{author}{\bibfnamefont{H.-M.} \bibnamefont{Eiter}},
  \bibinfo{author}{\bibfnamefont{M.}~\bibnamefont{Lavagnini}},
  \bibinfo{author}{\bibfnamefont{R.}~\bibnamefont{Hackl}},
  \bibinfo{author}{\bibfnamefont{E.~A.} \bibnamefont{Nowadnick}},
  \bibinfo{author}{\bibfnamefont{A.~F.} \bibnamefont{Kemper}},
  \bibinfo{author}{\bibfnamefont{T.~P.} \bibnamefont{Devereaux}},
  \bibinfo{author}{\bibfnamefont{J.-H.} \bibnamefont{Chu}},
  \bibinfo{author}{\bibfnamefont{J.~G.} \bibnamefont{Analytis}},
  \bibinfo{author}{\bibfnamefont{I.~R.} \bibnamefont{Fisher}},
  \bibnamefont{and} \bibinfo{author}{\bibfnamefont{L.}~\bibnamefont{Degiorgi}},
  \bibinfo{journal}{PNAS} \textbf{\bibinfo{volume}{110}}, \bibinfo{pages}{64}
  (\bibinfo{year}{2013}).

\bibitem[{\citenamefont{Moore et~al.}(2010)\citenamefont{Moore, Brouet, He, Lu,
  Ru, Chu, Fisher, and Shen}}]{Moore_2010}
\bibinfo{author}{\bibfnamefont{R.~G.} \bibnamefont{Moore}},
  \bibinfo{author}{\bibfnamefont{V.}~\bibnamefont{Brouet}},
  \bibinfo{author}{\bibfnamefont{R.}~\bibnamefont{He}},
  \bibinfo{author}{\bibfnamefont{D.~H.} \bibnamefont{Lu}},
  \bibinfo{author}{\bibfnamefont{N.}~\bibnamefont{Ru}},
  \bibinfo{author}{\bibfnamefont{J.-H.} \bibnamefont{Chu}},
  \bibinfo{author}{\bibfnamefont{I.~R.} \bibnamefont{Fisher}},
  \bibnamefont{and} \bibinfo{author}{\bibfnamefont{Z.-X.} \bibnamefont{Shen}},
  \bibinfo{journal}{Phys. Rev. B} \textbf{\bibinfo{volume}{81}},
  \bibinfo{pages}{073102} (\bibinfo{year}{2010}).

\bibitem[{\citenamefont{Gweon et~al.}(1998)\citenamefont{Gweon, Denlinger,
  Clack, Allen, Olson, DiMasi, Aronson, Foran, and Lee}}]{Gwe98}
\bibinfo{author}{\bibfnamefont{G.-H.} \bibnamefont{Gweon}},
  \bibinfo{author}{\bibfnamefont{J.~D.} \bibnamefont{Denlinger}},
  \bibinfo{author}{\bibfnamefont{J.~A.} \bibnamefont{Clack}},
  \bibinfo{author}{\bibfnamefont{J.~W.} \bibnamefont{Allen}},
  \bibinfo{author}{\bibfnamefont{C.~G.} \bibnamefont{Olson}},
  \bibinfo{author}{\bibfnamefont{E.~D.}~\bibnamefont{DiMasi}},
  \bibinfo{author}{\bibfnamefont{M.~C.} \bibnamefont{Aronson}},
  \bibinfo{author}{\bibfnamefont{B.}~\bibnamefont{Foran}}, \bibnamefont{and}
  \bibinfo{author}{\bibfnamefont{S.}~\bibnamefont{Lee}},
  \bibinfo{journal}{Phys. Rev. Lett.} \textbf{\bibinfo{volume}{81}},
  \bibinfo{pages}{886} (\bibinfo{year}{1998}).

\bibitem[{\citenamefont{Brouet et~al.}(2004)\citenamefont{Brouet, Yang, Zhou,
  Hussain, Ru, Shin, Fisher, and Shen}}]{Brouet_2004}
\bibinfo{author}{\bibfnamefont{V.}~\bibnamefont{Brouet}},
  \bibinfo{author}{\bibfnamefont{W.~L.} \bibnamefont{Yang}},
  \bibinfo{author}{\bibfnamefont{X.~J.} \bibnamefont{Zhou}},
  \bibinfo{author}{\bibfnamefont{Z.}~\bibnamefont{Hussain}},
  \bibinfo{author}{\bibfnamefont{N.}~\bibnamefont{Ru}},
  \bibinfo{author}{\bibfnamefont{K.~Y.} \bibnamefont{Shin}},
  \bibinfo{author}{\bibfnamefont{I.~R.} \bibnamefont{Fisher}},
  \bibnamefont{and} \bibinfo{author}{\bibfnamefont{Z.~X.} \bibnamefont{Shen}},
  \bibinfo{journal}{Phys. Rev. Lett.} \textbf{\bibinfo{volume}{93}},
  \bibinfo{pages}{126405} (\bibinfo{year}{2004}).

\bibitem[{\citenamefont{Komoda et~al.}(2004)\citenamefont{Komoda, Sato, Souma,
  Takahashi, Ito, and Suzuki}}]{Komoda_2004}
\bibinfo{author}{\bibfnamefont{H.}~\bibnamefont{Komoda}},
  \bibinfo{author}{\bibfnamefont{T.}~\bibnamefont{Sato}},
  \bibinfo{author}{\bibfnamefont{S.}~\bibnamefont{Souma}},
  \bibinfo{author}{\bibfnamefont{T.}~\bibnamefont{Takahashi}},
  \bibinfo{author}{\bibfnamefont{Y.}~\bibnamefont{Ito}}, \bibnamefont{and}
  \bibinfo{author}{\bibfnamefont{K.}~\bibnamefont{Suzuki}},
  \bibinfo{journal}{Phys. Rev. B} \textbf{\bibinfo{volume}{70}},
  \bibinfo{pages}{195101} (\bibinfo{year}{2004}).

\bibitem[{\citenamefont{Brouet et~al.}(2008)\citenamefont{Brouet, Yang, Zhou,
  Hussain, Moore, He, Lu, Shen, Laverock, Dugdale et~al.}}]{Brouet_2008}
\bibinfo{author}{\bibfnamefont{V.}~\bibnamefont{Brouet}},
  \bibinfo{author}{\bibfnamefont{W.~L.} \bibnamefont{Yang}},
  \bibinfo{author}{\bibfnamefont{X.~J.} \bibnamefont{Zhou}},
  \bibinfo{author}{\bibfnamefont{Z.}~\bibnamefont{Hussain}},
  \bibinfo{author}{\bibfnamefont{R.~G.} \bibnamefont{Moore}},
  \bibinfo{author}{\bibfnamefont{R.}~\bibnamefont{He}},
  \bibinfo{author}{\bibfnamefont{D.~H.} \bibnamefont{Lu}},
  \bibinfo{author}{\bibfnamefont{Z.-X.} \bibnamefont{Shen}},
  \bibinfo{author}{\bibfnamefont{J.}~\bibnamefont{Laverock}},
  \bibinfo{author}{\bibfnamefont{S.~B.} \bibnamefont{Dugdale}},
  \bibnamefont{et~al.}, \bibinfo{journal}{Phys. Rev. B}
  \textbf{\bibinfo{volume}{77}}, \bibinfo{pages}{235104}
  (\bibinfo{year}{2008}).

\bibitem[{\citenamefont{Lavagnini et~al.}(2008)\citenamefont{Lavagnini,
  Baldini, Sacchetti, Castroa, Delley, Monnier, Chu, Ru, Fisher, Postorino
  et~al.}}]{Lav08}
\bibinfo{author}{\bibfnamefont{M.}~\bibnamefont{Lavagnini}},
  \bibinfo{author}{\bibfnamefont{M.}~\bibnamefont{Baldini}},
  \bibinfo{author}{\bibfnamefont{A.}~\bibnamefont{Sacchetti}},
  \bibinfo{author}{\bibfnamefont{D.~D.} \bibnamefont{Castroa}},
  \bibinfo{author}{\bibfnamefont{B.}~\bibnamefont{Delley}},
  \bibinfo{author}{\bibfnamefont{R.}~\bibnamefont{Monnier}},
  \bibinfo{author}{\bibfnamefont{J.-H.} \bibnamefont{Chu}},
  \bibinfo{author}{\bibfnamefont{N.}~\bibnamefont{Ru}},
  \bibinfo{author}{\bibfnamefont{I.~R.} \bibnamefont{Fisher}},
  \bibinfo{author}{\bibfnamefont{P.}~\bibnamefont{Postorino}},
  \bibnamefont{et~al.}, \bibinfo{journal}{Phys. Rev. B}
  \textbf{\bibinfo{volume}{78}}, \bibinfo{pages}{201101(R)}
  (\bibinfo{year}{2008}).

\bibitem[{\citenamefont{Lavagnini et~al.}(2010)\citenamefont{Lavagnini, Eiter,
  Tassini, Muschler, Hackl, Monnier, Chu, Fisher, and Degiorgi}}]{Lav10}
\bibinfo{author}{\bibfnamefont{M.}~\bibnamefont{Lavagnini}},
  \bibinfo{author}{\bibfnamefont{H.-M.} \bibnamefont{Eiter}},
  \bibinfo{author}{\bibfnamefont{L.}~\bibnamefont{Tassini}},
  \bibinfo{author}{\bibfnamefont{B.}~\bibnamefont{Muschler}},
  \bibinfo{author}{\bibfnamefont{R.}~\bibnamefont{Hackl}},
  \bibinfo{author}{\bibfnamefont{R.}~\bibnamefont{Monnier}},
  \bibinfo{author}{\bibfnamefont{J.-H.} \bibnamefont{Chu}},
  \bibinfo{author}{\bibfnamefont{I.~R.} \bibnamefont{Fisher}},
  \bibnamefont{and} \bibinfo{author}{\bibfnamefont{L.}~\bibnamefont{Degiorgi}},
  \bibinfo{journal}{Phys. Rev. B} \textbf{\bibinfo{volume}{81}},
  \bibinfo{pages}{081101} (\bibinfo{year}{2010}).

\bibitem[{\citenamefont{Yusupov et~al.}(2008)\citenamefont{Yusupov, Mertelj,
  Chu, Fisher, and Mihailovic}}]{Yusopov_2008}
\bibinfo{author}{\bibfnamefont{R.~V.} \bibnamefont{Yusupov}},
  \bibinfo{author}{\bibfnamefont{T.}~\bibnamefont{Mertelj}},
  \bibinfo{author}{\bibfnamefont{J.-H.} \bibnamefont{Chu}},
  \bibinfo{author}{\bibfnamefont{I.~R.} \bibnamefont{Fisher}},
  \bibnamefont{and}
  \bibinfo{author}{\bibfnamefont{D.}~\bibnamefont{Mihailovic}},
  \bibinfo{journal}{Phys. Rev. Lett.} \textbf{\bibinfo{volume}{101}},
  \bibinfo{pages}{246402} (\bibinfo{year}{2008}).

\bibitem[{\citenamefont{Chen et~al.}(2014)\citenamefont{Chen, Hu, Dong, and
  Wang}}]{Che14}
\bibinfo{author}{\bibfnamefont{R.~Y.} \bibnamefont{Chen}},
  \bibinfo{author}{\bibfnamefont{B.~F.} \bibnamefont{Hu}},
  \bibinfo{author}{\bibfnamefont{T.}~\bibnamefont{Dong}}, \bibnamefont{and}
  \bibinfo{author}{\bibfnamefont{N.~L.} \bibnamefont{Wang}},
  \bibinfo{journal}{Phys. Rev. B} \textbf{\bibinfo{volume}{89}},
  \bibinfo{pages}{075114} (\bibinfo{year}{2014}).

\bibitem[{\citenamefont{Schmitt et~al.}(2011)\citenamefont{Schmitt, Kirchmann,
  Bovensiepen, Moore, Chu, Lu, Rettig, Wolf, Fisher, and Shen}}]{Schmitt_2011}
\bibinfo{author}{\bibfnamefont{F.}~\bibnamefont{Schmitt}},
  \bibinfo{author}{\bibfnamefont{P.~S.} \bibnamefont{Kirchmann}},
  \bibinfo{author}{\bibfnamefont{U.}~\bibnamefont{Bovensiepen}},
  \bibinfo{author}{\bibfnamefont{R.~G.} \bibnamefont{Moore}},
  \bibinfo{author}{\bibfnamefont{J.-H.} \bibnamefont{Chu}},
  \bibinfo{author}{\bibfnamefont{D.~H.} \bibnamefont{Lu}},
  \bibinfo{author}{\bibfnamefont{L.}~\bibnamefont{Rettig}},
  \bibinfo{author}{\bibfnamefont{M.}~\bibnamefont{Wolf}},
  \bibinfo{author}{\bibfnamefont{I.~R.} \bibnamefont{Fisher}},
  \bibnamefont{and} \bibinfo{author}{\bibfnamefont{Z.-X.} \bibnamefont{Shen}},
  \bibinfo{journal}{New J. of Physics} \textbf{\bibinfo{volume}{13}},
  \bibinfo{pages}{063022} (\bibinfo{year}{2011}).

\bibitem[{\citenamefont{Rettig et~al.}(2014)\citenamefont{Rettig, Chu, Fisher,
  Bovensiepen, and Wolf}}]{Rettig_2014}
\bibinfo{author}{\bibfnamefont{L.}~\bibnamefont{Rettig}},
  \bibinfo{author}{\bibfnamefont{J.-H.} \bibnamefont{Chu}},
  \bibinfo{author}{\bibfnamefont{I.~R.} \bibnamefont{Fisher}},
  \bibinfo{author}{\bibfnamefont{U.}~\bibnamefont{Bovensiepen}},
  \bibnamefont{and} \bibinfo{author}{\bibfnamefont{M.}~\bibnamefont{Wolf}},
  \bibinfo{journal}{Faraday Discuss.}   \textbf{\bibinfo{volume}{171}},  \bibinfo{pages}{299} (\bibinfo{year}{2014}).

\bibitem[{\citenamefont{Sobota et~al.}(2012)\citenamefont{Sobota, Yang,
  Analytis, Chen, Fisher, Kirchmann, and Shen}}]{Sobota_2012}
\bibinfo{author}{\bibfnamefont{J.~A.} \bibnamefont{Sobota}},
  \bibinfo{author}{\bibfnamefont{S.}~\bibnamefont{Yang}},
  \bibinfo{author}{\bibfnamefont{J.~G.} \bibnamefont{Analytis}},
  \bibinfo{author}{\bibfnamefont{Y.~L.} \bibnamefont{Chen}},
  \bibinfo{author}{\bibfnamefont{I.~R.} \bibnamefont{Fisher}},
  \bibinfo{author}{\bibfnamefont{P.~S.} \bibnamefont{Kirchmann}},
  \bibnamefont{and} \bibinfo{author}{\bibfnamefont{Z.-X.} \bibnamefont{Shen}},
  \bibinfo{journal}{Phys. Rev. Lett.} \textbf{\bibinfo{volume}{108}},
  \bibinfo{pages}{117403} (\bibinfo{year}{2012}).

\bibitem[{\citenamefont{Yang et~al.}(2014)\citenamefont{Yang, Sobota,
  Kirchmann, and Shen}}]{Yang_2014}
\bibinfo{author}{\bibfnamefont{S.-L.} \bibnamefont{Yang}},
  \bibinfo{author}{\bibfnamefont{J.~A.} \bibnamefont{Sobota}},
  \bibinfo{author}{\bibfnamefont{P.~S.} \bibnamefont{Kirchmann}},
  \bibnamefont{and} \bibinfo{author}{\bibfnamefont{Z.-X.} \bibnamefont{Shen}},
  \bibinfo{journal}{Applied Physics A} \textbf{\bibinfo{volume}{116}},
  \bibinfo{pages}{85} (\bibinfo{year}{2014}).

\bibitem[{\citenamefont{Giannozzi et~al.}(2009)}]{Giannozzi_2009}
\bibinfo{author}{\bibfnamefont{P.}~\bibnamefont{Giannozzi}},
  \bibinfo{journal}{J. Phys. Condens. Matter} \textbf{\bibinfo{volume}{21}},
  \bibinfo{pages}{395502} (\bibinfo{year}{2009}).

\bibitem[{\citenamefont{Perdew et~al.}(1996)\citenamefont{Perdew, Burke, and
  Ernzerhof}}]{Perdew_1996}
\bibinfo{author}{\bibfnamefont{J.~P.} \bibnamefont{Perdew}},
  \bibinfo{author}{\bibfnamefont{K.}~\bibnamefont{Burke}}, \bibnamefont{and}
  \bibinfo{author}{\bibfnamefont{M.}~\bibnamefont{Ernzerhof}},
  \bibinfo{journal}{Phys. Rev. Lett.} \textbf{\bibinfo{volume}{77}},
  \bibinfo{pages}{3865} (\bibinfo{year}{1996}).

\bibitem[{\citenamefont{{S}upplemental~{M}aterial (online)}()}]{sup_mat}
\bibinfo{author}{\bibnamefont{{S}upplemental~{M}aterial (online)}}.

\bibitem[{\citenamefont{Han et~al.}(2012)\citenamefont{Han, Tao, Mahanti,
  Chang, , Malliakas, Kanatzidis, and Ruan}}]{Han12}
\bibinfo{author}{\bibfnamefont{Tzong-Ru~T.} \bibnamefont{Han}},
  \bibinfo{author}{\bibfnamefont{Z.}~\bibnamefont{Tao}},
  \bibinfo{author}{\bibfnamefont{S.~D.} \bibnamefont{Mahanti}},
  \bibinfo{author}{\bibfnamefont{K.}~\bibnamefont{Chang}}, 
  \bibinfo{author}{\bibfnamefont{C.-Y.} \bibnamefont{Ruan}},
   \bibinfo{author}{\bibfnamefont{C.~D.} \bibnamefont{Malliakas}},
  \bibnamefont{and} \bibinfo{author}{\bibfnamefont{M.~G.} \bibnamefont{Kanatzidis}},
  \bibinfo{journal}{Phys. Rev. B} \textbf{\bibinfo{volume}{86}},
  \bibinfo{pages}{075145} (\bibinfo{year}{2012}).

\bibitem[{\citenamefont{Tao et~al.}(2013)\citenamefont{Tao, Han, and
  Ruan}}]{Tao13}
\bibinfo{author}{\bibfnamefont{Z.}~\bibnamefont{Tao}},
  \bibinfo{author}{\bibfnamefont{Tzong-Ru~T.} \bibnamefont{Han}},
  \bibnamefont{and} \bibinfo{author}{\bibfnamefont{C.-Y.} \bibnamefont{Ruan}},
  \bibinfo{journal}{Phys. Rev. B} \textbf{\bibinfo{volume}{87}},
  \bibinfo{pages}{235124} (\bibinfo{year}{2013}).

\bibitem[{\citenamefont{Khan and Allen}(1984)}]{Khan_1984}
\bibinfo{author}{\bibfnamefont{F.~S.} \bibnamefont{Khan}} \bibnamefont{and}
  \bibinfo{author}{\bibfnamefont{P.~B.} \bibnamefont{Allen}},
  \bibinfo{journal}{Phys. Rev. B} \textbf{\bibinfo{volume}{29}},
  \bibinfo{pages}{3341} (\bibinfo{year}{1984}).

\bibitem[{\citenamefont{Faure et~al.}(2013)\citenamefont{Faure, Mauchain,
  Papalazarou, Marsi, Boschetto, Timrov, Vast, Ohtsubo, Arnaud, and
  Perfetti}}]{Fauvre_2013}
\bibinfo{author}{\bibfnamefont{J.}~\bibnamefont{Faure}},
  \bibinfo{author}{\bibfnamefont{J.}~\bibnamefont{Mauchain}},
  \bibinfo{author}{\bibfnamefont{E.}~\bibnamefont{Papalazarou}},
  \bibinfo{author}{\bibfnamefont{M.}~\bibnamefont{Marsi}},
  \bibinfo{author}{\bibfnamefont{D.}~\bibnamefont{Boschetto}},
  \bibinfo{author}{\bibfnamefont{I.}~\bibnamefont{Timrov}},
  \bibinfo{author}{\bibfnamefont{N.}~\bibnamefont{Vast}},
  \bibinfo{author}{\bibfnamefont{Y.}~\bibnamefont{Ohtsubo}},
  \bibinfo{author}{\bibfnamefont{B.}~\bibnamefont{Arnaud}}, \bibnamefont{and}
  \bibinfo{author}{\bibfnamefont{L.}~\bibnamefont{Perfetti}},
  \bibinfo{journal}{Phys. Rev. B} \textbf{\bibinfo{volume}{88}},
  \bibinfo{pages}{075120} (\bibinfo{year}{2013}).

\bibitem[{\citenamefont{Tranquada et~al.}(1995)\citenamefont{Tranquada,
  Sternlieb, Axe, Nakamura, and Uchida}}]{Tranquada_1995}
\bibinfo{author}{\bibfnamefont{J.~M.} \bibnamefont{Tranquada}},
  \bibinfo{author}{\bibfnamefont{B.~J.} \bibnamefont{Sternlieb}},
  \bibinfo{author}{\bibfnamefont{J.~D.} \bibnamefont{Axe}},
  \bibinfo{author}{\bibfnamefont{Y.}~\bibnamefont{Nakamura}}, \bibnamefont{and}
  \bibinfo{author}{\bibfnamefont{S.}~\bibnamefont{Uchida}},
  \bibinfo{journal}{Nature} \textbf{\bibinfo{volume}{375}},
  \bibinfo{pages}{561} (\bibinfo{year}{1995}).

\bibitem[{\citenamefont{Yao et~al.}(2006)\citenamefont{Yao, Robertson, Kim, and
  Kivelson}}]{Yao_2006}
\bibinfo{author}{\bibfnamefont{H.}~\bibnamefont{Yao}},
  \bibinfo{author}{\bibfnamefont{J.~A.} \bibnamefont{Robertson}},
  \bibinfo{author}{\bibfnamefont{E.-A.} \bibnamefont{Kim}}, \bibnamefont{and}
  \bibinfo{author}{\bibfnamefont{S.~A.} \bibnamefont{Kivelson}},
  \bibinfo{journal}{Phys. Rev. B} \textbf{\bibinfo{volume}{74}},
  \bibinfo{pages}{245126} (\bibinfo{year}{2006}).

\end{thebibliography}
\end{document}